# Green's function for metamaterial superlens: Evanescent wave in the image

Wei Li[*] and XunYa Jiang[†]

*State Key Laboratory of Functional Materials for Informatics,*

*Shanghai Institute of Microsystem and Information Technology,*

*Chinese Academy of Sciences, Shanghai 200050, China*

We develop a new method to calculate the evanescent wave, the subdivided evanescent waves (SEWs), and the radiative wave, which can be obtained by separating the global field of the image of metamaterial superlens. The method is based on Green's function, and it can be applied in other linear systems. This study could help us to investigate the effect of evanescent wave on metamaterial superlens directly, and give us a new way to design new devices.

PACS number: 42.25.-p; 42.15.Eq

Keywords: metamaterial; superlens; evanescent wave; Green's function

[*] Email: waylee@mail.sim.ac.cn

[†] Email: xyjiang@mail.sim.ac.cn

Fax: +86-21-52419931

# I. INTRODUCTION

Early in 1968, Veselago [1] predicted that a new type of artificial metamaterial, which possesses simultaneously negative permittivity and permeability, could function as a lens to focus electromagnetic waves. This research direction was promoted by Pendry's work [2, 3] and other latter works [4–26]. They show that with such metamaterial lens, not only radiative waves but also evanescent waves, can be collected at its image, so the lens could be a superlens which can break through or overcome the diffraction limit of conventional imaging system. This beyond-limit property gives us a new window to design devices.

It is well-known that evanescent wave plays an important role in the beyond-limit property of the metamaterial superlens. Further more, evanescent wave becomes more and more important when the metamaterial devices enter sub-wavelength scales [27, 28]. Therefore, the quantitatively study of *pure* evanescent wave in the metamaterial superlens is very significant.

However, the quantitatively effects of *pure* evanescent wave in the metamaterial superlens have not been intensively studied, since so far almost all studies were only interested in the image properties with global field [8, 9], which is the summation of radiative wave and evanescent wave. On the other hand, many theoretical works were performed to study the metamaterial superlens, employing either finite-difference-time domain (FDTD) simulations [16] or some approximate approaches [29, 30].

However, one cannot obtain the rigorous pure evanescent wave by these numerical methods, since the image field of the metamaterial superlens obtained by FDTD is global field, and other approximate approaches cannot be rigorous.

In reviewing these existing efforts, we feel desirable to develop a rigorous method that can be used to study quantitatively the transient phenomena of the evanescent wave in the image of the metamaterial superlens. In this paper, we will present a new method based on the Green's function [7, 8] to serve this purpose. Our method can be successfully used to calculate the evanescent wave, as well as the radiative wave and the global field. The main idea of our method can be briefly illustrated as follows. Since the metamaterial superlens is a linear system, all dynamical processes can be solved by sum of multi-frequency components. And each frequency component can be solved by sum of multi-wavevector components. So we can use Green's function of multi-frequency components to obtain the strict numerical results. Therefore, our method based on the Green's function is strict, and it is quite a universe method in any linear system, for example, it can be used to study the two dimensional (2D) and three dimensional (3D) metamaterial superlens.

The content of the paper is as follows. We mainly focus on the details of the theory of our method in Section II. After that, in Section III, we will calculate the field of the image of a metamaterial superlens by our method, including evanescent wave, SEWs, and global field. And the method will be convinced by FDTD simulation. Finally, conclusions are presented in Section IV.

## II. THEORETICAL METHOD

In this section, we will focus on the theoretical details of our method. First, a time-dependent Green's function will be introduced. Then, based on such Green's function, the method how to obtain evanescent wave as well as radiative wave will be presented.

### A. A time-dependent Green's function

Here, a very useful time-dependent Green's function is introduced for the solution of inhomogeneous media. The time-dependent Green's function can be applied to study those dynamical scattering process[7, 22]. In the inhomogeneous media, the problem we study can be solved by Maxwell equations:

$$\nabla \times \vec{E}(r,t) = -\mu(r)\mu_0 \frac{\partial}{\partial t}\vec{H}(r,t),$$
$$\nabla \times \vec{H}(r,t) = \epsilon(r)\epsilon_0 \frac{\partial}{\partial t}\vec{E}(r,t) + \vec{J}(r,t), \quad (1)$$

where $\vec{J}(r,t) = \sigma(r,t)\vec{E}(r,t)$ is the current density and $\sigma(r,t)$ is the conductivity. We rewrite

Eq.(1) as:

$$\nabla \times \nabla \times \vec{E}(r,t) + \epsilon(r)\mu(r)\epsilon_0\mu_0\frac{\partial^2}{\partial t^2}\vec{E}(r,t) = -\mu(r)\mu_0\frac{\partial}{\partial t}[\sigma(r,t)\vec{E}(r,t)] \qquad (2)$$

To solve Eq.(2), we introduce a dynamic Green's function $\overset{\rightarrow\rightarrow}{G}(r,r';t,t')$, which satisfies:

$$\left(\nabla \times \nabla \times +\epsilon(r)\mu(r)\epsilon_0\mu_0\frac{\partial^2}{\partial t^2}\right)\overset{\rightarrow\rightarrow}{G}(r,r';t,t') = \delta(r-r')\delta(t-t')\overset{\rightarrow\rightarrow}{I} \qquad (3)$$

where $\overset{\rightarrow\rightarrow}{I}$ is a unit dyad. And in this system, when $r$ and $r'$ is given, $\overset{\rightarrow\rightarrow}{G}(r,r';t,t')$ is just a function of $(t-t')$ in time domain, so it yields

$$\overset{\rightarrow\rightarrow}{G}(r,r';t,t') = \overset{\rightarrow\rightarrow}{G}(r,r';t-t'). \qquad (4)$$

Therefore, Eq.(3) becomes:

$$\left(\nabla \times \nabla \times +\epsilon(r)\mu(r)\epsilon_0\mu_0\frac{\partial^2}{\partial t^2}\right)\overset{\rightarrow\rightarrow}{G}(r,r';t-t') = \delta(r-r')\delta(t-t')\overset{\rightarrow\rightarrow}{I} \qquad (5)$$

If $\overset{\rightarrow\rightarrow}{G}(r,r';t-t')$ is known, then the $\vec{E}$ field can be found as

$$\vec{E}(r,t) = -\mu_0 \int \overset{\rightarrow\rightarrow}{G}(r,r';t-t') \cdot \vec{J}(r',t')dr'dt' \qquad (6)$$

By Fourier transforming Eq.(5) from time domain to frequency domain, we obtain

$$\left(\nabla \times \nabla \times -\mu(r,\omega)\epsilon(r,\omega)\epsilon_0\mu_0\omega^2\right)\overset{\rightarrow\rightarrow}{G}(r,r';\omega) = \delta(r-r')\overset{\rightarrow\rightarrow}{I} \qquad (7)$$

where $\epsilon(r,\omega)$ and $\mu(r,\omega)$ are the relative permittivity and the relative permeability of the dispersive material at the frequency $\omega$, respectively, and $\overset{\rightarrow\rightarrow}{G}(r,r';\omega)$ is the Green's Function in frequency domain, which satisfies

$$\overset{\rightarrow\rightarrow}{G}(r,r';t-t') = \frac{1}{2\pi}\int d\omega\, \overset{\rightarrow\rightarrow}{G}(r,r';\omega)e^{-i\omega(t-t')} \qquad (8)$$

By Fourier transforming the electric field $\vec{E}(r,t)$ in time domain to the frequency domain, we obtain $\vec{E}(r,\omega)$, satisfying

$$\vec{E}(r,t) = \frac{1}{2\pi} \int d\omega e^{-i\omega t} \vec{E}(r,\omega) \tag{9}$$

Considering the general properties:

$$\begin{aligned}\epsilon(r,-\omega) &= \epsilon(r,\omega)^*, \\ \mu(r,-\omega) &= \mu(r,\omega)^*,\end{aligned} \tag{10}$$

and

$$\overrightarrow{\vec{G}}(r,r';-\omega) = \overrightarrow{\vec{G}}(r,r';\omega)^*. \tag{11}$$

Substituting Eq.(8), Eq.(9), Eq(10) and Eq.(11) into Eq.(6), we can obtain

$$\vec{E}(r,\omega) = \overrightarrow{\vec{G}}(r,r';\omega) \cdot \vec{E}(r',\omega). \tag{12}$$

Eq.(12) exhibits the role of $\overrightarrow{\vec{G}}(r,r';\omega)$ that $\overrightarrow{\vec{G}}(r,r';\omega)$ is a propagator for the field in the frequency domain between $r$ and $r'$. Here we have supposed that $\sigma(r',t) = 1$ in the source region, for simplicity. If $\vec{E}(r',\omega)$ is the exciting source's electric field in the frequency domain, i.e., $\vec{E}_s(r',\omega) = \vec{E}(r',\omega)$, then Eq.(12) can be rewritten as:

$$\vec{E}(r,\omega) = \overrightarrow{\vec{G}}(r,r';\omega) \cdot \vec{E}_s(r',\omega) \tag{13}$$

where $\vec{E}_s(r',\omega)$ is the spectrum of the exciting source.

So the field $\vec{E}(r,t)$ can be obtained by the inverse Fourier transformation:

$$\vec{E}(r,t) = \frac{1}{2\pi} \int d\omega e^{-i\omega t} \vec{E}(r,\omega), \tag{14}$$

where ω₀ is the working frequency of the exciting source.

Now the remaining problem is to solve Eq.(7) to get $\overrightarrow{\vec{G}}(r,r';\omega)$. With the exciting source $E_s(r',t)$ (whose spectrum is $E_s(r',\omega)$), after obtain $\overrightarrow{\vec{G}}(r,r';\omega)$, then we can calculate $\vec{E}(r,\omega)$ directly by Eq.(12).

As a typical example, the time-dependent Green's function for a three-layer inhomogeneous medium is presented, which is shown in Fig.1. The inhomogeneous media of the system can be described by:

$$\epsilon(r)\epsilon_0, \mu(r)\mu_0 = \begin{cases} \epsilon_0, \mu_0 & z > d \\ \epsilon_l^r \epsilon_0, \mu_l^r \mu_0 & 0 < z < d \\ \epsilon_0, \mu_0 & z < 0 \end{cases} \quad (15)$$

$\overset{\rightarrow\rightarrow}{G}(r, r'; \omega)$ is related to the transmission coefficient and/or reflection coefficient which are dependent on the boundary conditions of each layer. Here, we only consider the Green's function in the region that $z > d$. Under these conditions, $\overset{\rightarrow\rightarrow}{G}(r, r'; \omega)$ has been obtained in Ref.[7]. $\overset{\rightarrow\rightarrow}{G}(r, r'; \omega)$ is different in different spatial dimensions.

For 3D case, we can rewrite $\overset{\rightarrow\rightarrow}{G}(r, r'; \omega)$ as follows:

$$\overset{\rightarrow\rightarrow}{G}(r, r'; \omega) = -\frac{i\sigma(r', \omega)}{8\pi} \int \frac{1}{k_z} e^{-ik_z z} [T^{TE}(k_\parallel)(J_0(k_\parallel x) - J_2(k_\parallel x)) \\ + \frac{k_z^2}{k^2} T^{TM}(k_\parallel)(J_0(k_\parallel x) + J_2(k_\parallel x)) k_\parallel dk_\parallel], \quad (16)$$

where $J_0(z)$ and $J_2(z)$ are the usual zeroth-order Bessel function and second-order Bessel function, respectively, and $k_\parallel^2 + k_{lz}^2 = \epsilon_l^r \mu_l^r (\omega/c)^2$, $k_\parallel^2 + k_z^2 = (\omega/c)^2$ are the dispersion relation in the region $(0 < z < d)$ and the region $(z < 0, z > d)$ respectively. Here, $T^{TE}$ and $T^{TM}$ are the transmission coefficients for TE wave ($\vec{H}$ polarized) and TM wave ($\vec{E}$ polarized), respectively, which are respectively given by

$$T^{TE}(k_\parallel) = \frac{4\triangle e^{-ik_z d}}{(\triangle + 1)^2 e^{-ik_{lz} d} - (\triangle - 1)^2 e^{ik_{lz} d}}, \quad (17)$$

and

$$T^{TM}(k_\parallel) = \frac{4\triangle' e^{-ik_z d}}{(\triangle' + 1)^2 e^{-ik_{lz} d} - (\triangle' - 1)^2 e^{ik_{lz} d}}, \quad (18)$$

where $\triangle = \frac{k_{lz}}{k_z \mu_l^r}$ and $\triangle' = \frac{k_{lz}}{k_z \epsilon_l^r}$.

$\sigma(r', \omega)$ is the conductivity in source region, i.e., $\vec{J}(r', \omega) = \sigma(r', \omega)\vec{E}_s(r', \omega)$. For simplicity, $\sigma(r', \omega)$ could be assumed as an arbitrary constant independent $\omega$, here we set $\sigma(r', \omega) = 1$. So Eq.(16) becomes:

$$\overset{\leftrightarrow}{G}(r, r'; \omega) = -\frac{i}{8\pi} \int \frac{1}{k_z} e^{-ik_z z} [T^{TE}(k_\parallel)(J_0(k_\parallel x) - J_2(k_\parallel x)) + \frac{k_z^2}{k^2} T^{TM}(k_\parallel)(J_0(k_\parallel x) + J_2(k_\parallel x))k_\parallel dk_\parallel], \quad (19)$$

For 2D case, the wave vector $k_\parallel = k_x$, we have

$$\overset{\leftrightarrow}{G}(r, r'; \omega) = -\frac{i}{4\pi} \int \frac{e^{ik_x x}}{k_z} T^{TE}(k_x) e^{ik_z z} dk_x. \quad (20)$$

And for one-dimension, we have

$$\overset{\leftrightarrow}{G}(r, r'; \omega) = -\frac{i}{2k} T^{TE}(0) e^{-ikz}. \quad (21)$$

After $\overset{\leftrightarrow}{G}(r, r'; \omega)$ is obtained, we can obtain the field in the frequency domain in the region $z > d$ by Eq.(12). And then by the inverse Fourier transformation, the field in time domain can be obtained.

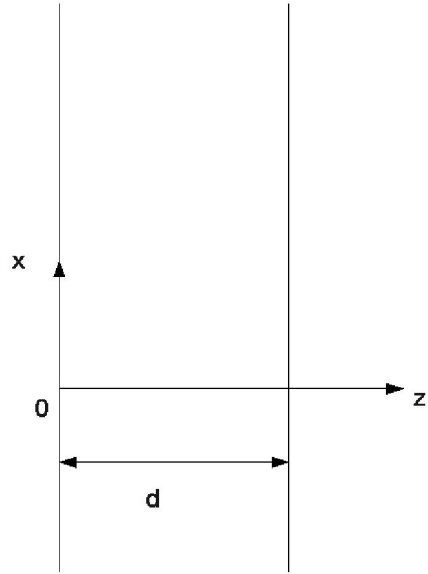

Fig. 1. The shematic of the three-layer inhomogeneous medium.

**B. The Green's function for radiative wave and evanescent wave**

Now, we will apply a time-dependent Green's function for the radiative wave and evanescent wave. This Green's function is directly developed from the Green's function introduced in the Sec. IIA. The schematic model is shown in Fig.1. As we know, the plane solution wave for the electric field in vacuum is of the form $E_z(r_{||}, z, t) = E_{z0}\exp(i(k_{||}r_{||} + k_z z - \omega t))$, where $k_{||}$ and $k_z$ are wave numbers along the $xy$ plane and $z$ directions respectively, and they satisfy the dispersion relation as follows: $k_{||}^2 + k_z^2 = \omega^2/c^2$, where $c$ is the light velocity in the vacuum. In the case of $k_{||}^2 < \omega^2/c^2$, $k_z$ is real, corresponding the radiative wave along the $z$ direction. While if $k_{||}^2 > \omega^2/c^2$, $k_z$ is imaginary, corresponding the evanescent wave along the $z$ direction. Similarly, in the slab region, the real or imaginary $k_{lz}$ corresponds to the radiative wave or the evanescent wave along $z$ direction, respectively.

For simplicity, we first consider the 2D system, in which $k_{||} = k_x$. The global field of image region is the superposition of radiative wave and evanescent wave. We can rewrite Eq.(20), Eq.(13) and Eq.(14) as follows:

$$\vec{\vec{G}}(r, r'; \omega; k_{min}, k_{max}) = -\frac{i}{4\pi} \int_{k_{min}}^{k_{max}} \frac{e^{ik_x x}}{k_z} T^{TE}(k_x) e^{ik_z z} dk_x, \qquad (22)$$

$$\vec{E}(r; \omega; k_{min}, k_{max}) = \vec{\vec{G}}(r, r'; \omega; k_{min}, k_{max}) \cdot \vec{E}_s(r', \omega), \qquad (23)$$

and

$$\vec{E}(r; t; k_{min}, k_{max}) = \frac{1}{2\pi} \int d\omega e^{-i\omega t} \vec{E}(r; \omega; k_{min}, k_{max}), \qquad (24)$$

where $k_{min} < k_x < k_{max}$ is the integral range. The integral range is of great significance, since different integral range corresponds to different wave. For example, the integral range $[k_{min} \to -\infty, k_{max} \to \infty]$ is for global field, and the range $[k_{min} = -\omega/c, k_{max} = \omega/c]$ is for radiative wave.

From Eq.(22) to Eq.(24), we can directly calculate the radiative wave and evanescent wave. In the case of radiative wave ($k_x^2 < \omega^2/c^2$), the integral range is $[k_{min} = -\omega/c, k_{max} = \omega/c]$, so the radiative wave Green's function $\overset{\rightarrow\rightarrow}{G}_{rad}(r,r';\omega)$ satisfies

$$\overset{\rightarrow\rightarrow}{G}_{rad}(r,r';\omega) = \overset{\rightarrow\rightarrow}{G}(r,r';\omega; k_{min}=-\omega/c, k_{max}=\omega/c)$$
$$= -\frac{i}{4\pi}\int_{-\omega/c}^{\omega/c}\frac{e^{ik_x x}}{k_z}T^{TE}(k_x)e^{ik_z z}dk_x \tag{25}$$

for radiative wave.

In the case of evanescent wave ($k_x^2 > \omega^2/c^2$), the integral range is $k_x > |\omega/c|$, so the evanescent wave Green's function $\overset{\rightarrow\rightarrow}{G}_{eva}(r,r';\omega)$ satisfies:

$$\overset{\rightarrow\rightarrow}{G}_{eva}(r,r';\omega) = 2\lim_{k_{max}\to\infty}\overset{\rightarrow\rightarrow}{G}(r,r';\omega; k_{min}=\omega/c, k_{max})$$
$$= -\frac{i}{2\pi}\lim_{k_{max}\to\infty}\int_{\omega/c}^{k_{max}}\frac{e^{ik_x x}}{k_z}T^{TE}(k_x)e^{ik_z z}dk_x \tag{26}$$

for evanescent wave.

And for the global field, the global field Green's function $\overset{\rightarrow\rightarrow}{G}_{glob}(r,r';\omega)$ satisfies:

$$\overset{\rightarrow\rightarrow}{G}_{glob}(r,r';\omega) = \overset{\rightarrow\rightarrow}{G}_{rad}(r,r';\omega) + \overset{\rightarrow\rightarrow}{G}_{eva}(r,r';\omega) \tag{27}$$

Additionally, we can also focus our observation on the *subdivided evanescent wave* (SEW), with a certain integral range $[k_{min}=k_x^a, k_{max}=k_x^b]$. The SEW, with a certain integral range, can also be obtained from Eq.(22) to Eq.(24) such that

$$\overset{\rightarrow\rightarrow}{G}_{SEW}(r,r';\omega)|_{k_{min}=k_x^a}^{k_{max}=k_x^b} = \overset{\rightarrow\rightarrow}{G}(r,r'; k_x^a, k_x^b)$$
$$= -\frac{i}{4\pi}\int_{k_x^a}^{k_x^b}\frac{e^{ik_x x}}{k_z}T^{TE}(k_x)e^{ik_z z}dk_x. \tag{28}$$

As an typical example, a SEW with an integral range $[k_{min}=1.1\omega/c, k_{max}=1.2\omega/c]$, whose Green's function can be obtained easily from Eq.(28) as:

$$\overset{\rightarrow\rightarrow}{G}_{SEW}(r,r';\omega)|_{k_{min}=1.1\omega/c}^{k_{max}=1.2\omega/c} = -\frac{i}{4\pi}\int_{1.1\omega/c}^{1.2\omega/c}\frac{e^{ik_x x}}{k_z}T^{TE}(k_x)e^{ik_z z}dk_x.$$

In this way, we can obtain the Green's function for the SEW with any integral range. Obviously,

evanescent wave could be regarded as the superposition of SEWs. Therefore, we have

$$\vec{\vec{G}}_{eva}(r,r';\omega) = \sum_{SEWs} \vec{\vec{G}}_{SEW}(r,r';\omega). \tag{29}$$

From Eq.(28), Eq.(26), Eq.(25), and Eq.(27), one can obtain the SEW Green's function, one can obtain the SEW Green's function $\vec{\vec{G}}_{SEW}(r,r';\omega)$, the evanescent wave Green's function $\vec{\vec{G}}_{eva}(r,r';\omega)$, the radiative wave Green's function $\vec{\vec{G}}_{rad}(r,r';\omega)$, and the global field Green's function $\vec{\vec{G}}_{glob}(r,r';\omega)$, respectively. Substituting them to Eq.(23) and Eq.(24) respectively, we can obtain the field of the SEW, the evanescent wave, the radiative wave, and the global field, respectively.

For the 3D system, obviously, the methods to get Green's function for the SEW, the evanescent wave, the radiative wave, and the global field are respectively very similar with the above discussion, i.e., just replace $k_x$ by $k_\parallel$ and let $k_\parallel^2 = k_x^2 + k_y^2$ in Eq.(28), Eq.(26), Eq.(25), and Eq.(27), respectively.

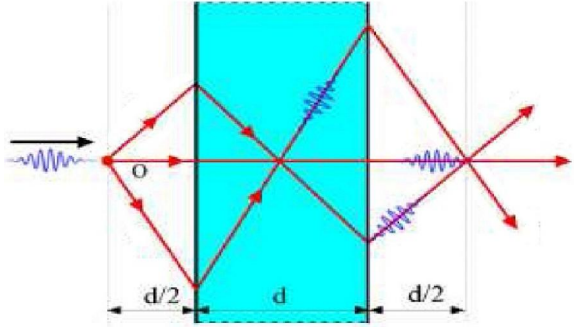

Fig. 2. The schematic diagram of our 2D model.

### III. NUMERICAL RESULTS

In order to show the applications of our method, we will discuss the 2D metamaterial superlens, which is shown in Fig.2. The thickness of the metamaterial slab is $d$, which is placed at the $xy\text{-}plane$ between $z = d/2$ and $z = 3d/2$. The source is set in the object plane at $(z = 0)$. Obviously, the image will be formed in the image plane at $z = 2d$. The source is the quasi-monochromatic random source with the field expressed as $E_s(r,t) = U_s(r,t)exp(-i\omega_0 t)$, where $U_s(r,t)$ is a slow-varying random function, $\omega_0 = 1.33 \times 10^{15}/s$ is the central frequency of our random source (the details of the random source can be seen in Ref.[29]). The exciting source $E_s(r, t)$ is a quasi-monochromatic field with the central frequency $\omega_0$, whose electric field $\vec{E}_s(t)$ and frequency spectrum $\vec{E}_s(\omega)$ are shown in Fig.3(a) and (b), respectively. In this paper, only the TE modes are investigated (the TE modes have the magnetic field perpendicular to the two-dimensional plane of our model).

The inhomogeneous media of the metamaterial superlens system are described by:

$$\epsilon(r), \mu(r) = \begin{cases} \epsilon_0, \mu_0 & z > 3/2d \\ \epsilon_0 \epsilon_l^r, \mu_0 \mu_l^r & 1/2d < z < 3/2d \\ \epsilon_0, \mu_0 & z < 1/2d \end{cases} \qquad (30)$$

The negative relative permittivity $\epsilon_l^r$ and the negative relative permeability $\mu_l^r$ of metamaterial are phenomenologically introduced by the Lorenz model, satisfying as follows:

$$\epsilon_l^r(\omega) = \mu_l^r(\omega) = 1 + \omega_p^2/(\omega_a^2 - \omega^2 - i\Delta\omega \cdot \omega) \qquad (31)$$

where $\omega_a = 1.884 \times 10^{15}/s$ and $\Delta\omega = 1.88 \times 10^{14}/s$ are the resonant frequency and the resonant linewidth of the "resonators" in the metamaterial respectively, and $\omega_p = 10 \times \omega_a$ is the plasma frequency. At $\omega = \omega_0$, we have $\epsilon_l^r = \mu_l^r = -1.0 + i0.0029$.

In order to excite the evanescent wave strong enough in the image of the metamaterial superlens, the distance $d/2$ between the source and the superlens should be small enough. Here we choose $d = \lambda_0/2$, where $\lambda_0 = 1.42\mu m$ is the wavelength corresponding to the central frequency $\omega_0$.

For this metamaterial superlens system, it is very easy to obtain $\overleftrightarrow{G}_{glob}(r, r'; \omega)$, $\overleftrightarrow{G}_{eva}(r, r'; \omega)$, and $\overleftrightarrow{G}_{SEW}(r, r'; \omega)$ from Eq.(26) to Eq.(28), respectively. Let $r = 2d$ and $r' = 0$, and thus the Green's functions for the image will be obtained. After that, we can obtain the global field, the evanescent wave and the SEW of the image via:

$$\vec{E}_{glob}(2d, t) = \frac{1}{2\pi} \int d\omega e^{-i\omega t} \vec{E}_{glob}(2d, \omega)$$

$$\vec{E}_{eva}(2d, t) = \frac{1}{2\pi} \int d\omega e^{-i\omega t} \vec{E}_{eva}(2d, \omega) \qquad (32)$$

$$\vec{E}_{SEW}(2d, t) = \frac{1}{2\pi} \int d\omega e^{-i\omega t} \vec{E}_{SEW}(2d, \omega)$$

respectively, where

$$\vec{E}_{glob}(2d, \omega) = \vec{\vec{G}}_{glob}(2d, 0; \omega) \cdot \vec{E}_s(\omega)$$

$$\vec{E}_{eva}(2d, \omega) = \vec{\vec{G}}_{eva}(2d, 0; \omega) \cdot \vec{E}_s(\omega) \qquad (33)$$

$$\vec{E}_{SEW}(2d, \omega) = \vec{\vec{G}}_{SEW}(2d, 0; \omega) \cdot \vec{E}_s(\omega).$$

The numerical results calculated by our method are shown in Fig.3. Fig.3(c) (up, the blue one), (d) and (e) show the global field, the evanescent wave, and two typical SEWs respectively. The integral $k_x$ range of the two SEWs are $[k_{min} = 1.1\omega/c, k_{max} = 1.2\omega/c]$ (shown in Fig.3(e) (up, the black one)), and $[k_{min} = 1.3\omega/c, k_{max} = 1.4\omega/c]$ (shown in Fig.3 (e) (down, the blue one)), respectively.

In order to convince our method, FDTD simulation is also applied to calculate the field of the image, which is shown in Fig.3(c) (down, the green one). Comparison with the results calculated by our method and FDTD, we can see they coincide with each other very well. In addition, we also calculate the frequency spectrum of the image by our method, as shown in Fig.3(b) (down, the red one). Comparing the spectra of source and image, we can find they are very close to each other. This result also agrees with the Ref.[29]. Therefore, our method is convincible, which can be used to obtain the pure evanescent wave, the SEW, and the global field effectively.

### III. CONCLUSION

In conclusion, we have numerically and theoretically obtained the evanescent wave, as well as the SEWs, separating from the global field, which is based on the Green's function. This study could help us to investigate the effect of evanescent wave on metamaterial superlens directly, and give us a new way to design new devices.

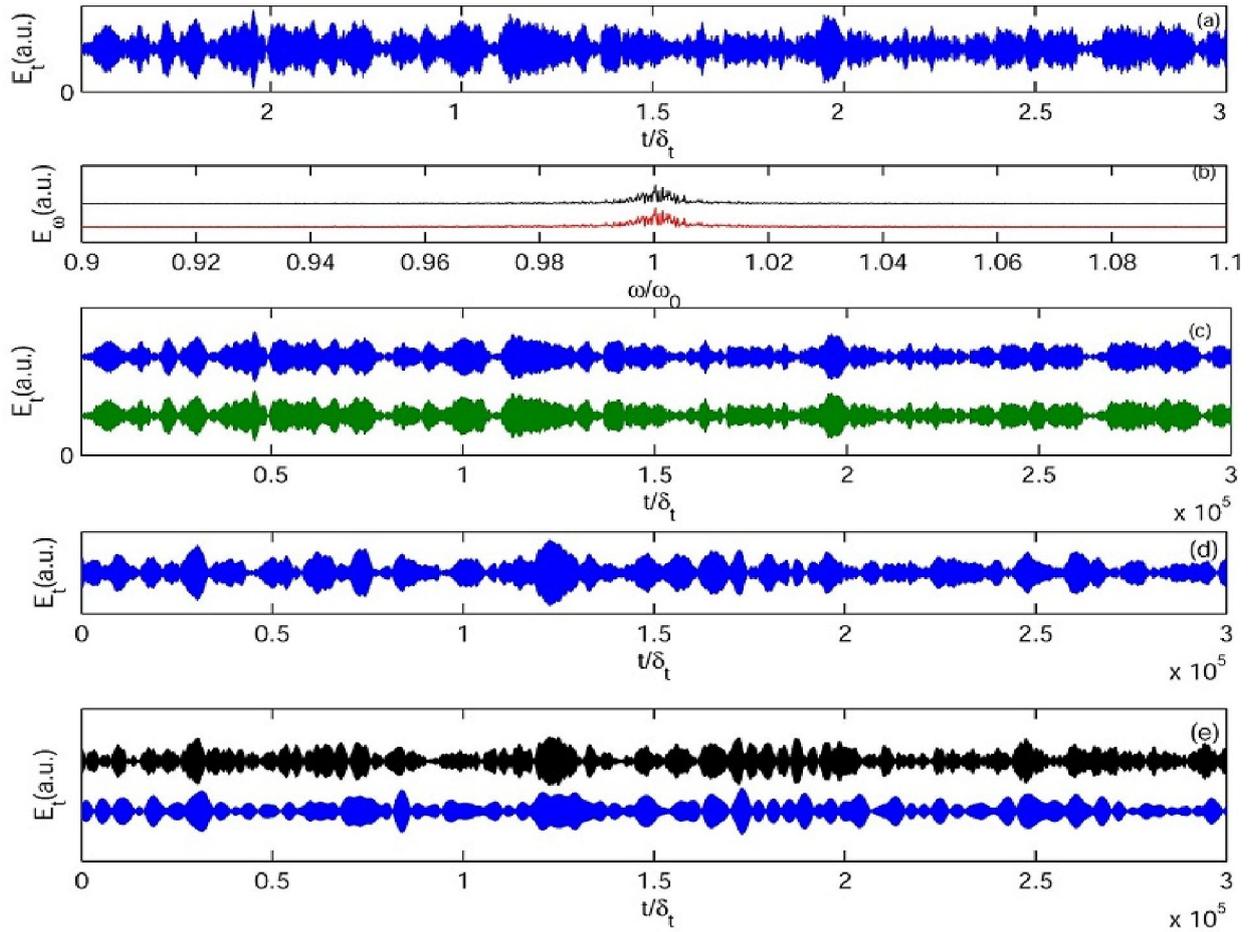

Fig. 3. (a)The electric field of the source. (b) The spectrum of the source (up) and the image obtained by our method (down) in unit of $\omega_0$. (c) The electric field of image vs time. The global field calculated by our method (up) and by FDTD (down). (d) The evanescent wave of image calculated by our method (up). (e) Two typical SEWs of the image.

## ACKNOWLEDGEMENT

This work is supported by the NSFC (Grant Nos. 11174309, 11004212, 10704080, 60877067, and 60938004) and the STCSM (Grant No. 11ZR1443800).

## REREFENCES